\documentclass[aps,12pt,nofootinbib]{revtex4}

\usepackage{graphicx,graphics}
\usepackage{enumerate}
\usepackage{subfigure}
\usepackage{amsmath}
\usepackage{amsfonts}
\usepackage{amssymb}
\usepackage{amsthm}
\usepackage{psfrag}

\catcode`\@=11
\def\numberbysection{\@addtoreset{equation}{section}
\def\theequation{\arabic{section}.\arabic{equation}}}
\numberbysection

\def\a{\alpha}

\def\Z{\mathbb{Z}}

\begin{document}

\title{A Loop Reversibility and Subdiffusion of the Rotor-Router Walk}

\author{Vl.V. Papoyan$^1$, V.S. Poghosyan$^2$ and V.B. Priezzhev$^1$}
\affiliation{
$^1$Bogoliubov Laboratory of Theoretical Physics,\\ Joint Institute for Nuclear Research, 141980 Dubna, Russia\\
$^2$Institute for Informatics and Automation Problems\\ NAS of Armenia, 0014 Yerevan, Armenia
}

\begin{abstract}
The rotor-router model on a graph describes a discrete-time walk accompanied by the deterministic evolution of configurations of rotors randomly placed on vertices of the graph. We prove the following property:
if at some moment of time, the rotors form a closed clockwise contour on the planar graph,
then the clockwise rotations of rotors generate a walk which enters into the contour at some vertex $v$,
performs a number of steps inside the contour so that the contour formed by rotors becomes anti-clockwise,
and then leaves the contour at the same vertex $v$.
This property generalizes the previously proved theorem for the case when the rotor configuration inside the contour is a cycle-rooted spanning tree, and all rotors inside the contour perform a full rotation. We use the proven property for an analysis of the sub-diffusive behavior of the rotor-router walk.
\end{abstract}

\maketitle


\section{Introduction}

The Eulerian walkers model introduced in \cite{PDDK} as an example of self-organized criticality \cite{BTW,Dhar},
was rediscovered later by researchers in different fields and attracted much attention due to its simple algorithmic structure
\cite{CS,LP05,LP07,LP08,HP,AngelHol,HussSava,FGLP}.
Some properties of the Eulerian walkers, stated in \cite{PDDK} and in the related paper \cite{PPS},
were formulated then as rigorous graph-theoretical theorems \cite{HLMPPW}.
In the mathematical literature, the model received the name ``rotor-router walk'' proposed by Jim Propp \cite{CS}, who
considered it as a derandomized analogue of the ordinary random walk.

\begin{figure}[!ht]
\includegraphics[width=160mm]{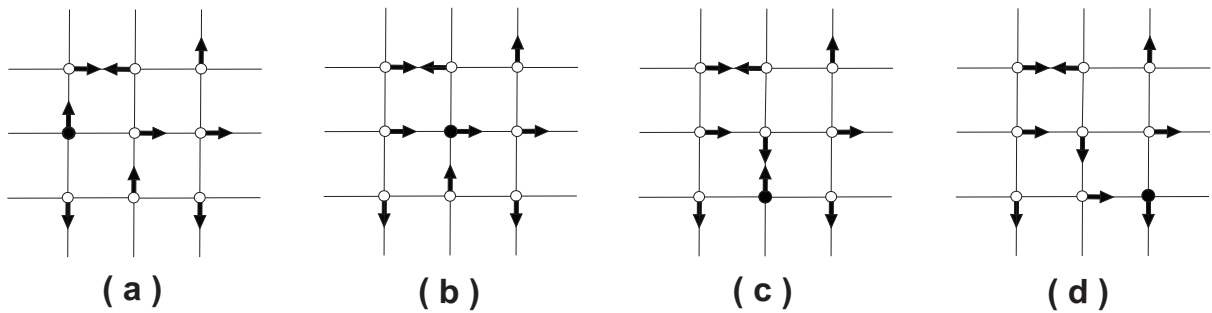}
\caption{Circles denote the lattice sites. (a) The chip is originally in the filled circle where the arrow is directed "up".
(b) The chip rotates the arrow clockwise and moves right. (c) The next clockwise rotation sends the chip down. (d) The last position
of the chip is the right lower corner.}
\label{steps}
\end{figure}

A formal definition of the rotor-router walk is given in Section II.
In a less formal way, one considers a lattice with arrows attached to the lattice sites.
The arrow at every site is directed to one of its neighbors on the lattice.
A particle (chip) performs a walk jumping from a site to a neighboring site.
Arriving to a given site, the chip changes the direction of the arrow at this site to next position, for instance 90 degrees clockwise, and moves to the neighbor pointed by new position of the arrow.
Thus, given an initial orientation of arrows on the whole lattice, the rotor-router walk is fully deterministic.
Fig.\ref{steps} illustrates three steps of the rotor walk on the square lattice.

At present, the basic properties of the rotor-router walks on finite graphs can be considered as well established. The walk starting from an arbitrary vertex of a finite Eulerian graph $G$ (see the definition below) eventually settles into an Eulerian circuit where each directed edge of $G$ is visited exactly once. When the walker is in the Eulerian circuit, configurations of rotors associated to vertices of $G$ are recurrent. A striking property of the recurrent state is a filling of loops formed by rotors on a planar graph $G$ under clockwise rotations of rotors: if a sequence of rotors makes up a closed clockwise oriented contour $C$ on $G$ at some moment of the Eulerian circuit, the walker enters into contour $C$ in a vertex $v\in C$, performs a full rotation of each rotor internal to $C$, changes the orientation of rotors on $C$ to opposite and leaves the contour in the same vertex $v$ \cite{PPS,HLMPPW}. This property of the contour is called {\it reversibility}. Recently, Chan, Church and Grochow \cite{CCG} have noticed that the reversibility of all cycles on a graph is a test whether the given graph is planar or not. Specifically, they proved that a connected  loopless graph is planar if and only if all cycles are reversible under the rotor-router dynamics.

The behavior of the rotor-router walk on infinite graphs is quite different.
Let $G$ be an infinite square lattice with the rotors associated to each vertex.
Each rotor has four possible directions ordered clockwise.
The initial orientations of rotors are independently distributed with equal probabilities for each direction.
The walker starting at the origin performs a diffusion-like motion.
However, the exponent $\nu$ in the law of the mean-square displacement for $n$ steps
$\langle r^2(n)\rangle \sim n^{2\nu}$ differs considerably from its random walk value $\nu=1/2$.
Extensive computer experiments show that $\nu=1/3$ with high precision.
A qualitative derivation of the subdiffusion law $\nu=1/3$ in \cite{PDDK} is far from a rigorous proof although
the arguments given there seem to be credible.
These arguments are based on an assumption that the walker's trajectory covers a kernel of the cluster almost densely before reaching the shell of the cluster again.
Then, the radius of kernel grows with a velocity inversely proportional to the area of the kernel, so the average radius $R\sim n^{1/3}$.
Despite a strong numerical evidence for the assumption, there are no theoretical explanations of this unusual behavior of the rotor-router walk.

The aim of the present work is to consider the motion of the rotor-router walk in a random environment in more detail.
Analyzing the trajectories of a walk inside closed contours, we find a rigorous rule of visiting a closed contour emerged in a random rotor configuration.
Namely, we prove a property which we call the {\it weak reversibility} for the rotor-router walks, where the condition of the full rotation of each rotor internal to the contour is released.
Instead, some rotors inside $C$ perform a partial rotation or do not move at all. Nevertheless, two properties of reversibility are still retained: the rotor-router walk entering $C$ in a vertex $v \in C$ leaves the contour in the same vertex $v$, and the clockwise orientation of rotors on the contour $C$ becomes anti-clockwise.

We consider the rotor-router walk inside the contour and we describe precisely which part of the contour interior is filled by the trajectory of the walk.
Then, we discuss a possible relation of the loop reversibility to the subdiffusion of the rotor-router walk on the infinite lattice.
We observe that the set of vertices, where rotors make up clockwise contours, grows according to a definite rule, namely, the sequence of such vertices generates a spiral-like structure. Moreover, the obtained spirals, being random, in average obey an unexpected law which we call the asymptotical {\it Archimedean} property. The spiral-like structure of the reference points of the closed clockwise contours together with the weak reversibility of contours provides the subdiffusion behavior of the rotor-router walk.

\section{The model and a basic theorem}

Consider a directed graph (digraph) $G=(V,E)$ with a set of vertices $V=V(G)$ and a set of directed edges $E=E(G)$.
We assume that there are no self-loops or multiple edges in $G$ although the definition of the model can be extended
to this case. If for each edge directed from $v$ to $w$ there exists an edge directed from $w$ to $v$, we call the graph $G$ bidirected. A bidirected graph can be obtained by replacing each edge of an undirected graph with a pair of directed edges, one in each direction. A spanning subgraph $G\,'$ of a bidirected graph $G$ is a digraph with the set of vertices $V(G\,') = V(G)$ and a set of edges $E(G\,') \subseteq E(G)$.

A path of length $n$ from vertex $a \in V$ to $b \in V$ is a sequence of distinct vertices
$v_1,v_2, \dots, v_{n+1}$ such that $v_i$ and $v_{i+1}$ are connected by an edge $e_{i} \in E$, $i=1,2,\dots,n$, $v_1=a$, $v_{n+1}=b$.
The path becomes a cycle if $a=b$.
A shortest possible cycle has length $2$ and consists of two adjacent vertices $v_1$, $v_2$,
which are connected by a pair of edges from $v_1$ to $v_2$ and back.
We call such cycles dimers by analogy with lattice dimers covering two neighboring vertices.
A cycle formed by more than two edges is called contour.

An Eulerian circuit on a finite digraph is a walk which starts and ends on the same vertex and visits each directed edge exactly once.
If such a walk exists, the digraph is called Eulerian.
A digraph is strongly connected if for any two distinct vertices $v$, $w$ there are paths from $v$ to $w$ and from $w$ to $v$.
A strongly connected digraph $G=(V,E)$ is Eulerian if and only if for each vertex $v \in V$ in-degree and out-degree of $v$ are equal.
In particular, the one-component bidirected graph is Eulerian.

The rotor-router model is defined as follows. Consider an arbitrary connected digraph $G = (V,E)$.
Denote the number of outgoing edges (out-degree) from the vertex $v \in V$ by $d_v$.
The total number of edges of $G$ is $|E| = \sum_{v\in V} d_v$.
Each vertex $v \in V$ is associated with a rotor, which is directed along one of the outgoing edges from $v$.
The rotor directions at the vertex $v$ are specified by an integer variable $\a_v$,
which takes values from $0 \leq \a_v \leq d_v - 1$ for $d_v \geq 1$.

The set $\rho \equiv \{ \a_v|\; v\in V,\; 0 \leq \a_v \leq d_v - 1\}$ defines the rotor configuration.
Starting with an arbitrary rotor configuration, one drops a chip to a vertex of $G$ chosen at random.
At each time step the chip arriving at a vertex $v$, first changes the rotor direction from $\a_v$ to $\a_v + 1$\,,
and then moves one step along the new rotor direction from $v$ to the corresponding neighboring vertex.
The periodicity of $\alpha_v$ is assumed $(\alpha_v \pm d_v \equiv \alpha_v)$.

The rotor configuration $\rho$ can be considered as a spanning subgraph of $G$ $(\rho \subset G)$
with the set of vertices $V(\rho)=V(G)$ and the set of directed edges $E(\rho) \subset E(G)$ coinciding with the rotors.
The state of the system at any moment of time is given by the pair $(\rho, v)$ of the rotor configuration $\rho$ and the position of the chip $v \in V$.

A vertex $v \in V$ is called sink if its out-degree $d_v = 0$.
In the absence of sinks, i.e. when each vertex has at least one outgoing edge, the motion of the chip does not stop.
If iterating the rotor-router operation from the state $(\rho,v)$ eventually leads back to $(\rho,v)$ we say that $(\rho,v)$ is recurrent; transient otherwise.
According to arguments in \cite{PDDK}, the rotor-router walk, started from an arbitrary initial state $(\rho, v)$ on a finite graph, passes transient states and enters into a recurrent state continuing the motion in the limiting cycle which is the Eulerian circuit of the graph.

In \cite{U} a useful notion of {\it unicycle} is introduced (see also \cite{RetProb}).
A connected spanning subgraph of a digraph $G$, in which every vertex has one outgoing edge contains exactly one cycle.
The state $(\rho, v)$ is called unicycle if the set of edges $E(\rho)$ contains a unique directed cycle and $v$ lies on this cycle.
Then, two basic properties of the rotor-router model on the Eulerian graphs can be formulated in terms of unicycles.

{\it Property A} (\cite{HLMPPW}, Theorem 3.8). Let $G$ be a strongly connected digraph.
Then a single-chip-rotor state $(\rho, v)$ on $G$ is recurrent if and only if it is a unicycle.

The rotor states that are not unicycles, are transient.
In contrast to recurrent states, they appear at the initial stage of evolution
up to the moment when the system enters into the Eulerian circuit.

{\it Property B} (\cite{HLMPPW}, Lemma 4.9). Let $G$ be an Eulerian digraph with $m$ edges. Let $(\rho, v)$ be
a unicycle in $G$. If one iterates the rotor-router operation $m$ times starting from $(\rho, v)$, the chip traverses
an Euler tour of $G$, each rotor makes one full turn, and the state of the system returns to $(\rho, v)$.

A theorem on reversibility of loops at the recurrent state \cite{PPS,HLMPPW} mentioned in Introduction reads :

{\it Theorem 1} (\cite{PPS} and \cite{HLMPPW}, Corollary 4.11).
Let $G$ be a bidirected planar graph with the outgoing edges at each vertex ordered clockwise.
Let $(\rho, v)$ be a unicycle on $G$ with the cycle $C$ oriented clockwise.
After the rotor-router walk makes some number of steps, each rotor internal to $C$ has performed a full rotation,
each rotor external to $C$ has not moved, and each rotor on $C$ has performed a partial rotation so that $C$ is now oriented anti-clockwise.

Our aim in the next section is a proof of the theorem on weak reversibility.

\section{Contours in random rotor environment}

The theorem on weak reversibility that we are going to prove, is related to planar digraphs.
However, we start with a lemma which is valid for general directed planar and non-planar graphs.

\begin{figure}[!ht]
\includegraphics[width=80mm]{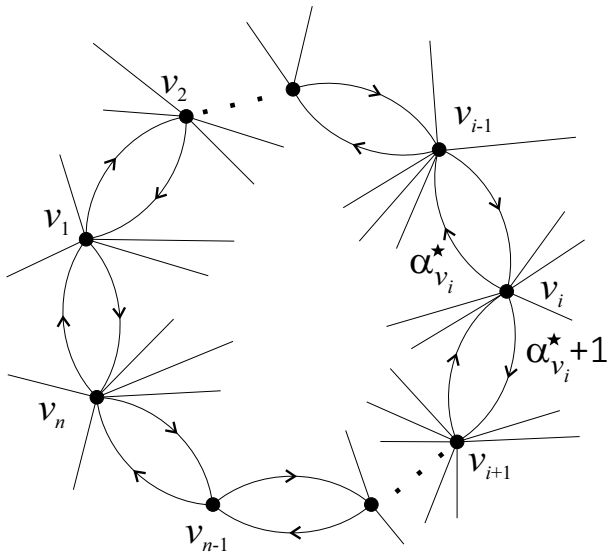}
\caption{ Bidirected contour obeying domino ordering. }
\label{domino}
\end{figure}

Consider a bidirected contour $C = (v_1,v_2, \dots, v_{n})$ in a digraph $G=(V,E)$,
that is a contour in which the vertices $v_i$ and $v_{i+1}$ are connected by two edges $e^{+}_{i}, e^{-}_{i+1} \in E$
one in each direction, $1 \leq i \leq n$, where the periodicity of the indices is assumed
($v_{i \pm n} \equiv v_{i}$, $e^{+}_{i \pm n} \equiv e^{+}_{i}$, $e^{-}_{i \pm n} \equiv e^{-}_{i}$).
Given the rotor-router model defined on $G$, we say that the bidirected contour $C$ obeys the {\it domino ordering}
if for each rotor at $v_i$, $1 \leq i \leq n$ there exists a direction $\alpha^{\star}_{v_i}$ such that
the rotor $\alpha^{\star}_{v_i}$ points from $v_i$ to $v_{i-1}$ and $\alpha^{\star}_{v_i} + 1$ points from $v_i$ to $v_{i+1}$.
The directions $\alpha^{\star}_{v_1}, \ldots, \alpha^{\star}_{v_n}$ are called negative with respect to $C$, whereas the directions
$\alpha^{\star}_{v_1}+1, \ldots, \alpha^{\star}_{v_n}+1$ are called positive, correspondingly (Fig.\ref{domino}).

{\it Lemma 1.} Given an arbitrary finite Eulerian digraph $G$, let $C=(v_1,\dots,v_n)$ be a bidirected contour obeying domino ordering.
Let the rotor-router walk starts at the vertex $v_n$ from an initial rotor configuration with positive directions of all rotors at $v_i$ ($i=1,2,\ldots,n)$.
Then, after some number of steps, the walk produces a configuration with negative directions
$\alpha^{\star}_{v_1},\dots,\alpha^{\star}_{v_n}$.
The moments $t_i$ $(i=1,2,\ldots,n)$, when the directions $\alpha^{\star}_{v_i}$ are reached,
are ordered as follows: $0 < t_n < t_{n-1} < \cdots < t_2 < t_1 \leq |E|$.

{\it Proof. } Since the digraph $G$ is finite, each vertex $v \in C$ is visited by the walk indefinitely many times.
Let $T$ be the first moment of time when the rotor at $v_n$ returns to the initial direction $\alpha^{\star}_{v_n} + 1$.
None of the edges of $G$ is visited twice by the walk  before $T$.
Indeed, assume the first repeated visit happens for an edge $e\in E$ directed from a vertex $v\neq v_n$.
Then, vertex $v$ is visited more than $d_v$ times, i.e. more than once from one of its neighbors $v^{'}\neq v$,
that contradicts to the assumption that the first edge, visited twice, is the edge directed from vertex $v$.
Each rotor being in direction $\alpha^{\star}_v + 1$ at some moment of time $t$ will change its direction to $\alpha^{\star}_v$
if and only if the vertex $v$ is visited by the walk exactly $d_v - 1$ times.
The first vertex $v$ of the contour $C$ being visited $d_v-1$ times, is the vertex $v_n$,
since other vertices $v_i$ of $C$ are not visited neither from $v_{i+1}$ nor from $v_{i-1}$ before, and there are no repeated visits from any vertices. The moment when the direction of arrow at $v_n$ becomes $\alpha^{\star}_{v_n}$ is $t_n < T$.
The next vertex is $v_{n-1}$, as it has been already visited from $v_n$. Direction $\alpha^{\star}_{v_{n-1}}$ is reached at moment $t_{n-1}$,where $t_{n}<t_{n-1}<T$ because the arrow at $v_n$ cannot return to the initial direction $\alpha^{\star}_{v_{n}}+1$ earlier than the arrow at $v_1$ is in direction $\alpha^{\star}_{v_1}$.
The process continues for vertices $v_{n-2},v_{n-3},\dots,v_1$.
Non of rotors at $v_i$, $i=1,\ldots,n$ has direction $\alpha^{\star}_{i} + 1$ before $t_1 < T\leq |E|$,
since the vertex $v_i$ is not yet visited from $v_{i-1}$ $\Box$.

If $(\rho,v)$ is a unicycle on the planar graph $G$ with the contour $C$, then the internal subgraph $G_{int}\subset G$
formed by rotors inside $C$ is a spanning forest, i.e. the graph whose vertices coincide with internal vertices of $C$ and edges form trees rooted at $C$. Now, consider the situation when the rotors at the internal vertices of $C$ have arbitrary orientations.
Since each internal vertex contains a rotor, the internal subgraph $G_{int}$ remains a spanning subgraph.
The edges of $G_{int}$ can be grouped into disjoint cycles and trees rooted at these cycles.
The trees which are not rooted at the internal cycles, have roots at the external contour $C$.

All contours can be either oriented clockwise or anti-clockwise.
By an analogy with the unicycle, we introduce the {\it multicycle}
as a graph containing exactly $k$ cycles together with $k$ chips at vertices $a_0,a_1,\dots,a_{k-1}$ belonging to the cycles.
For multicycles, we will use the notation $(\rho,a_0,a_1,\dots,a_{k-1})$.

\begin{figure}[!ht]
\includegraphics[width=60mm]{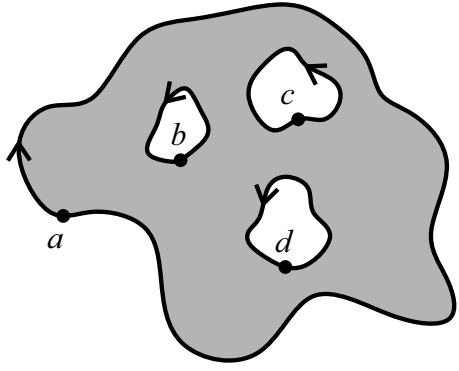}
\caption{Multicycle: external clockwise contour containing the vertex $a$ and three anti-clockwise contours containing vertices $b,c,d$.  }
\label{multicycle}
\end{figure}

First, we consider a situation where the external contour is clockwise and all internal contours are anti-clockwise (Fig.\ref{multicycle}).
We prove a theorem on weak reversibility generalizing the Theorem 1.

{\it Theorem 2.} Let $G$ be a connected bidirected planar graph and $(\rho,a_0,a_1,\dots,a_{k-1})$ be a multicycle with the external contour $C_0$ oriented clockwise together with $k-1$ internal cycles $C_1,\dots,C_{k-1}$ oriented anti-clockwise.
The rotor-router operation is sequentially applied to the chip at $a_0 \in C_0$ until the moment when the chip returns to $a_0$, and the rotor
at $a_0$ is made oriented anticlockwise.
Then, the same is applied to chips at $a_1,\dots,a_{k-1}$ until the moments when chips starting from $a_i \in C_i$ return to $a_i$ and the rotors at $a_i$ are made oriented clockwise.
Then, all rotors on $C_0$ are becoming oriented anticlockwise, whereas all rotors on $C_1,\dots,C_{k-1}$ become oriented clockwise, also all vertices internal to $C_0$ and external to $C_1,\dots,C_{k-1}$ perform a full rotation.

\begin{figure}[!ht]
\includegraphics[width=60mm]{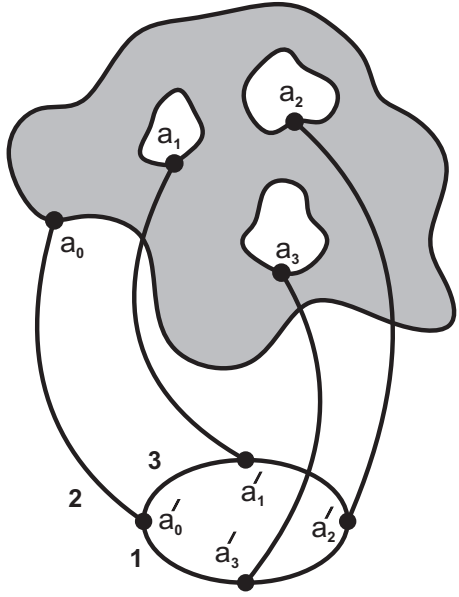}
\caption{Unicycle constructed from the external clockwise contour and three anti-clockwise contours.
Numbers 1,2,3 at $a_0^{'}$ correspond to the order of bonds adjacent to this vertex. }
\label{unicycle}
\end{figure}

{\it Proof.} Given a multicycle $(\rho,a_0,a_1,\dots,a_{k-1})$, we construct an auxiliary graph $G^{\star}$
performing the following operations (Fig.\ref{unicycle}):

1. reverse all the clockwise rotors on the external contour $C_0$ to anticlockwise, also all anticlockwise rotors on internal cycles
$C_1,\dots,C_{k-1}$ to clockwise;

2. remove all the vertices and edges of $G$ external to $C_0$, also remove the vertices and edges internal to any of $C_i,\ i=1,\dots,k-1$;

3. add an additional contour $C_0^{\,'}$ consisting of $k$ vertices $a_0^{'},a_1^{'},\dots,a_{k-1}^{'}$.

4. connect vertices $a_i$ and $a_i^{'}$ by edges for all $i=0,\dots,k-1$;

5. change the directions of rotors at $a_0,a_1,\dots,a_{k-1}$ so that new directions are toward $a_0^{'},a_1^{'},\dots,a_{k-1}^{'}$; the order of directions of rotors at
$a_0^{'},a_1^{'},\dots,a_{k-1}^{'}$ is shown in Fig.\ref{unicycle} for the vertex $a_0^{'}$.

6. put a single chip at $a_0^{'}$.

The obtained single-chip-and-rotor state on the auxiliary graph $G^{\star}$ is the unicycle $U^{\star}$.
After these preparations, we can use the property B for the unicycle $U^{\star}$.
According to this property, the chip starting at $a_0^{'}$ returns to $a_0^{'}$,
whereas the unicycle $U^{\star}$ returns to its initial state.
The initial state of rotors in $U^{\star}$ coincides with the final state of rotors at vertices of $C_0,\dots,C_{k-1}$ and also at vertices internal to $C_0$ and external to $C_1,\dots,C_{k-1}$ claimed in the theorem.

Thus, we have only to prove that, during their evolution, these rotors reach the initial state indicated in the theorem.
The additional contour $C_0^{\,'}$ provides the walker successive visits of the vertices $a_0,a_1,\dots,a_{k-1}$ from $a_0^{'},a_1^{'},\dots,a_{k-1}^{'}$.
It is straightforward to see that the first visit of the rotor walk to $C_0$ is followed by $|C_0|$ steps which make the orientation of rotors clockwise,
whereas the first visit to every internal contour $C_i$ is followed by $|C_i|$ steps which make the orientation on $C_i$ anticlockwise.
Since the sequences of $|C_0|$ steps for the contour $C_0$ and $|C_i|$ steps for contours $C_i$ both are deterministic, we can ignore all of them,
replacing each sequence by reversed orientations of contours just before the first visit to $C_0$ and $C_i, i=1,\dots,k-1$.
Thus, the conditions of the theorem are fulfilled. $\Box$

{\it Corollary.} Let $v_1,v_2, \dots, v_{n}$ be vertices of $C_0$ ordered clockwise, and $v_n = a_0$ be the first vertex of $C_0$
visited by the rotor-router walk. Then the moments of time $t_1,\dots,t_n$, when the rotors at $v_1,v_2, \dots, v_{n}$
become anticlockwise with respect to $C_0$, are ordered as follows: $ t_n < t_{n-1} < \cdots < t_2 < t_1$.

{\it Proof.} After operations 1 and 2, listed in the proof of Theorem 2, the bidirected contour $C_0$ obeys domino ordering.
Then, the statement of corollary follows from Lemma 1. $\Box$

The condition of anticlockwise orientation of the internal cycles $C_1,\dots,C_{k-1}$ in Theorem 2 can be released.
Indeed, let $C_j, 1\leq j \leq n-1$ be a cycle oriented clockwise.
If there are no internal cycles inside $C_j$, then the first hit of a chip to the contour $C_j$ converts, by Theorem 1,
the orientation of $C_j$ into anticlockwise after visiting its interior.
Then, we can simply skip the interval of the chip evolution between the moments of entry and
exit from $C_j$ considering  the orientation of $C_j$ as anticlockwise.
If the interior of cycle $C_j$ contains anticlockwise cycles only, we can skip the mentioned interval of the evolution, by Theorem 2.
If there are clockwise cycles inside $C_j$, then we can apply the skipping procedure recursively.
Since each generation of cycles is enclosed in the previous one, we eventually reach a smallest cycle which does not contain any cycle inside.

\section{Weak reversibility and the subdiffusion law}

As mentioned in the introduction, the crucial concepts used for explanation of the subdiffusion behavior of the rotor-router walk
on the infinite lattice are the kernel-shell structure of the cluster of visited vertices and the assumption
that the trajectory of the walk covers the kernel almost densely in intervals between periodic returns to the shell.
This property was approved by extensive simulations \cite{PDDK}, but yet a detailed explanation has not been presented.
In this section, we use Theorem 2 on weak reversibility to show how the rotor-router dynamics leads to the subdiffusion law.

Let $G=(V,E)$ be an infinite square lattice and $\rho_0$ be the initial rotor configuration with $\{\alpha_v|v\in V\}$ taken
uniformly from the set $\{0,1,2,3\}$.
The rotor-router walk starts the motion from the origin and performs $T$ steps forming a cluster of visited vertices and edges.
We fix all moments of time when the rotors form clockwise contours, and numerate these contours as they appear.
Assume that after $T_k^{(in)}$ steps, the rotor-chip configuration creates a clockwise contour $C_k$.
The $T_k^{(in)}$-th step is directed to the vertex $v_k \in C_k$.
According to Theorem 2, the chip returns to $v_k$ after visiting the interior of $C_k$,
and the clockwise contour becomes anticlockwise.
We denote by $T_k^{(out)}$ the moment of exit from $C_k$ and put a label $s_k$ at $v_k$.
Continuing, we obtain a sequence of contours $C_1,C_2,\dots$ and labels $s_1,s_2,\dots$.
We are interested in relative disposition of contours and labels appeared during time $T$.

To distinguish between rotors which did not move until the given moment of time and those involved into the motion,
we call the moved rotors {\it activated}.
A current position of all activated rotors, except the last one, forms a single rooted tree $\mathbb{T}$.
The location of the root is a current position of the chip.
If the position coincides with the label $s_k$, the root $s_k$ of the tree $\mathbb{T}$
is connected with the origin by a radial branch $0 \rightarrow s_k$.
For $t>T_k^{(out)}$, one possibility is that the diffusing chip reaches a vertex connected with the branch  $0 \rightarrow s_k$
by a sequence of activated or non-activated rotors so that the next appeared clockwise contour is adjacent to $C_k$.
Otherwise, the chip creates one or several isolated clockwise contours having no common edge with the branch $0 \rightarrow s_k$, one by one.
When, eventually, the chip reaches a vertex connected with the branch  $0 \rightarrow s_k$,
the next clockwise contour adjacent to $C_k$ appears, and all isolated contours (which have already changed their orientation
from clockwise to anti-clockwise by this time) become connected with $C_k$.

\begin{figure}[!ht]
\includegraphics[width=90mm]{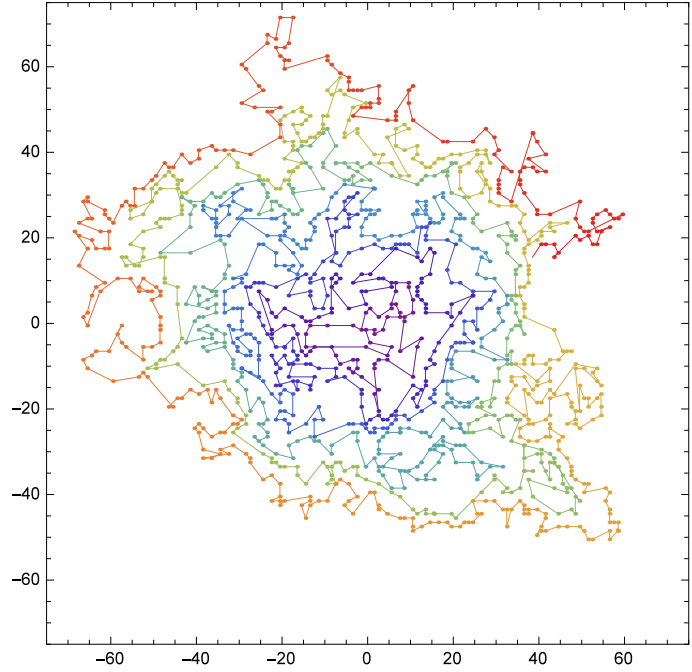}
\caption{The spiral-like sequence of labels in a cluster obtained after $T=10^5$ steps of the rotor-router walk.
Rainbow colors from violet to red correspond to increasing time.}
\label{labels}
\end{figure}

Consider a label $s_k$ situated near the boundary of the cluster of activated rotors.
A preferable position for $s_{k+1}$ is on the right side of the branch $0 \rightarrow s_k$ to provide clockwise orientation of the contour
$C_{k+1}$ if it has common edges with $0 \rightarrow s_k$.
Then, the preferable direction of successive positions of labels $s_k,s_{k+1},s_{k+2},\dots$ is clockwise with respect to the origin of the cluster.
Since the size of cluster grows with time, the positions $s_k,s_{k+1},s_{k+2},\dots$ form a spiral-like structure (Fig. \ref{labels}).

A spiral is called Archimedean if, in polar coordinates $(r,\theta)$, it can be described by the equation
\begin{equation}
r=a+b\,\theta,
\label{Archimed}
\end{equation}
where $a$ and $b$ are real numbers.

If a spiral is random, we will say that it is asymptotically Archimedean in average if
\begin{equation}
 \frac{\langle r \rangle}{\langle \theta\rangle}\rightarrow b \ \ \ \text{for}\ \ \  \theta \rightarrow \infty,
\label{definition}
\end{equation}
where the average is taken over the uniformly distributed  states of the spiral.
Above definitions of the sequence of contours $C_1,\dots,C_n$ and labels $s_1,\dots, s_n$ allow us to formulate a key conjecture:

{\it Conjecture 1.}  Given a random configuration of rotors on the infinite square lattice,
the spiral of labels $s_1,s_2,s_3,\dots$ generated by the rotor-router walk is asymptotically Archimedean in average
taken over the uniformly distributed initial states of rotors.

A numerical verification of the conjecture is not a trivial task.
The simulation of $10^4$ walks of length $T=10^9$ shows that
the ratio $r(s_n)/\theta(s_n)$ as a function of $n$ for a single spiral fluctuates around a constant value up to very large $n$.
However, for the ratio of the root-mean-square values, we estimated the convergence to the constant as:
\begin{equation}
\frac{\surd \overline{\langle r(s_n)^2 \rangle }}{\surd \overline{\langle \theta(s_n)^2 \rangle }} \simeq 1.85 + O(n^{-\frac{1}{4}}).
\label{conjecture1}
\end{equation}

Now, we can represent the rotor-router walk as a sequence of transitions between labels $s_1 \rightarrow s_2 \rightarrow \dots \rightarrow s_{n}$ and the time intervals $[T_1^{(in)},T_1^{(out)}]$, $[T_2^{(in)},T_2^{(out)}],\dots$ which the chip spends for visiting the interiors $A(C_1),A(C_2),\dots$ of contours $C_1,C_2, \dots$.
For each $k\geq 1$, there exists $k^{'}$ such that the walk between $s_k$ and $s_{k^{'}}$ is a loop of the spiral.
Consider a domain $A_{k,k^{'}}$ constituted of contours associated with labels
$s_k,s_{k+1},\dots,s_k^{'}$,
\begin{equation}
A_{k,k^{'}}=\bigcup_{j=k}^{k^{'}} A(C_j)
\label{domain}
\end{equation}
The area of $A_{k,k^{'}}$ is of an order $R^2$, where $R$ is the loop radius.
By Theorem 2, the rotor-router walk visits interiors of all contours in $A_{k,k^{'}}$ one by one and,
therefore, the total time $\Delta T$ needed for closing the loop is of an order $R^2$.
An advance $\Delta R$ of the loop radius for the time interval $\Delta T$ is of the order of the spiral step $2\pi b$.
Thus, the velocity of growth of $R$ is proportional to the inverse area of the loop,
\begin{equation}
\frac{\Delta R}{\Delta T} \sim \frac{1}{R^2},
\label{equation}
\end{equation}
from which we obtain the asymptotical law $R(T) \sim T^{1/3}$ for large $T$.

The arguments leading to (\ref{equation}) are qualitative, as well as those in \cite{PDDK}.
However, now we are equipped with three new concepts, namely: the weakly reversible contours,
the labels marking the clockwise contours and the spiral-like structure of labels.
They make the statements of \cite{PDDK} more constructive and show clear meaning of the
shell-kernel structure of the growing clusters used for the explanation of the subdiffusion law.

We can conclude that the statistics of labels appears to be a basic feature of the rotor-router walk,
which determines its long range behavior.
Beside the spiral-like structure, the labels can be characterized by the average time interval between them.
Specifically, the average number of steps $\delta t_n = T_{n+1}^{(in)} - T_{n}^{(out)}$ between labels $s_n$ and $s_{n+1}$,
i.e. the interval between exiting from the clockwise contour $C_n$ and entering to $C_{n+1}$,
tends to a constant value for large $n$:
\begin{equation}
\langle  \delta t_n \rangle \simeq 6.81 + O(n^{-\frac{1}{2}}).
\label{conjecture2}
\end{equation}
Since the spiral step tends to the constant, the spacial average density
of the labels $\rho(r)$ does not depend on distance $r$ from the origin in the large interval of $r$, $ 1 \ll r \ll T^{1/3}$.
Our Monte-Carlo simulations confirm this property.
We found that near the origin, for which the spiral structure is not pronounced yet,
the average density has a peak with $\rho(0) \simeq 0.37$, which decays rapidly to the plateau value $\rho_{st} \simeq 0.13$.
For $r > T^{1/3}$, the density vanishes sharply in accordance with the subdiffusion law $R(T)\sim T^{1/3}$.

\section*{Acknowledgments}
This work was supported by the JINR Program ``Smorodinsky -- Ter-Antonyan'', and
the State Committee of Science MES RA, in frame of the research project No. SCS 13-1B170.


\begin{thebibliography}{99}


\bibitem{PDDK}
V.B. Priezzhev, D. Dhar, A. Dhar, and S. Krishnamurthy.
Eulerian walkers as a model of self-organized criticality.
{\em Phys.\ Rev.\ Lett.} 77:5079--5082 (1996).

\bibitem{BTW}
P. Bak, C. Tang, and K. Wiesenfeld. Self-organized criticality: an explanation of the $1/f$ noise.
\newblock {\em Phys.\ Rev.\ Lett.} 59(4), 381--384 (1987).

\bibitem{Dhar}
D. Dhar. Self-organized critical state of sandpile automaton models.
{\em Phys.\ Rev.\ Lett.} 64(14), 1613--1616 (1990).

\bibitem{CS}
J.N. Cooper and J. Spencer. Simulating a random walk with constant error.
{\em Combin.\ Probab.\ Comput.} 15(6), 815--822 (2006).

\bibitem{LP05}
L. Levine and Y. Peres. The rotor-router shape is spherical.
{\em Math.\ Intelligencer} 27(3), 9--11 (2005).

\bibitem{LP07}
L. Levine and Y. Peres.
Strong spherical asymptotics for rotor-router aggregation and the divisible sandpile.
Potential Analysis 2009, Volume 30, Issue 1, pp 1--27.

\bibitem{LP08}
L. Levine and Y. Peres.
Spherical asymptotics for the rotor-router model in $\Z^d$.
{\em Indiana Univ. Math. J.} 57, 431--450 (2008).

\bibitem{HP}
A.E. Holroyd and J. Propp. Rotor walks and Markov Chains.
Algorithmic Probability and Combinatorics, volume 520 of Contemporary Mathematics, 105--126 (2010).

\bibitem{AngelHol} O.Angel, A.E. Holroyd. Recurrent Rotor-Routed Configurations. J. Comb. 3(2), 185--194 (2012).

\bibitem{HussSava} W. Huss, E. Sava. Transience and recurrebce of rotor-router walks on directed covers of graphs.
Electron. Commun. Probab. 17 (2012), no. 41, 1--13.

\bibitem{FGLP}
L. Florescu, S. Ganguly, L. Levine and Y. Peres.
Escape rates for rotor walk in $\mathbb{Z}^d$.
SIAM J. Discrete Math. 28(1), 323--334 (2014).

\bibitem{PPS}
A.M. Povolotsky, V.B. Priezzhev, and R.R. Shcherbakov.
Dynamics of Eulerian walkers.
{\em Phys.\ Rev.\ E} 58, 5449--5454 (1998).

\bibitem{HLMPPW}
A.E. Holroyd, L. Levine, K. Meszaros, Y. Peres, J. Propp and D.B. Wilson.
Chip-Firing and Rotor-Routing on Directed Graphs.
{\em Progress in Probability} 60, 331--364 (2008).

\bibitem{CCG} M.Chan, T.Church, and J.A. Grochow. Rotor-routing and spanning trees on planar graphs.
Int Math Res Notices (2015) 2015(11): 3225--3244.

\bibitem{U}
L. Levine and Y. Peres.
The looping constant of $\Z^d$. {\em Random Struct. Alg.} 45(1), 1--13 (2014)

\bibitem{RetProb}
V.S. Poghosyan, V.B. Priezzhev and P. Ruelle.
Return probability for the loop-erased random walk and mean height in the Abelian sandpile model: a proof.
{\em J. Stat. Mech.:Theor.Exp.} (2011) P10004.



\end{thebibliography}
\end{document}